\documentclass[runningheads]{llncs}
\usepackage{graphicx}
\usepackage{multirow}
\usepackage{rotating}
\usepackage{array}
\usepackage{subcaption}

\begin{document}
\newcommand{\etal}{\textit{et al}. }

\newcommand{\comment}[1]{}

\title{Query Intent Detection from the SEO Perspective}

\author{Samin Mohammadi, Mathieu Chapon, Arthur Fr\'emond }
\institute{
Search ForeSight, 68 rue Marjolin
92300, Levallois Perret, France \\
\email{samin.mohammadi, mathieu.chapon, arthur.fremond@search-foresight.com}
}

\maketitle

\section{Abstract}
\label{0_abstract}
Google users have different intents from their queries such as acquiring information, buying products, comparing or simulating services, looking for products and so on. Understanding the right intention of users helps to provide i) better content on web pages from the Search Engine Optimization (SEO\footnote{Search engine optimization is the process of increasing the quality and quantity of website traffic by increasing the visibility of a website\cite{SEOwiki}}) perspective and ii) more user-satisfying results from the search engine perspective. In this study, we aim to identify the user query's intent by taking advantage of Google results and machine learning methods. Our proposed approach is a clustering model that exploits some features to detect query's intent. A list of keywords extracted from the clustered queries is used to identify the intent of a new given query. Comparing the clustering results with the intents predicted by filtered keywords show the efficiency of the extracted keywords for detecting intents. 

\section{Introduction}
\label{sec:intro}
Search engines try to predict users' intentions from their queries to provide the most accurate results. In the concept of search engines, intent detection generally is a classification problem aiming to find the intention of input data which is mainly in the format of text. Intent detection has different applications in Natural Language Processing (NLP) tasks such as question answering, chatbots, and search engines. In the tasks like question answering, the intent is normally one or several words selected from the question. While in the context of search engines, user queries are mainly divided into three groups in terms of their intent including \textit{Informational}, \textit{Navigational}, and \textit{Transactional}. Informational queries are the most searched ones where the user's goal is looking for certain information by asking questions or searching for the keywords, for example ``who is the CEO of Apple?''. The user's intent from the navigational queries is to redirect to a specific website, for example ``Apple''. Transactional queries intend to do a transaction such as a purchase, for example ``buy an iPhone''. 

Understanding the right intent of the user's query will help search engines to present the most related results which finally leads to higher user satisfaction. Plenty of researches have been done to identify query's intent \cite{zhou2017survey} \cite{jansen2007determining}. Different types of models are proposed using machine learning \cite{ashkan2009classifying} \cite{jansen2007determining} and natural language processing (NLP) models \cite{pirvu2018predicting} \cite{meng2017dialogue}. 

In this study, we look at the intent detection problem from the perspective of SEO which is different than a search engine's perspective. The main goal here is to identify the user's intent from what a search engine, such as Google, provides to users in response to their query. Identifying user's intent from this point of view will help the SEO to suggest better content for the websites. The suggested content will be aligned with what the search engine expects for a given query. It is finally speeding up and facilitating the semantic analysis from the SEO processing. The model's outputs are some set of keywords to filter the queries and label their intent easier and faster.
The proposed model consists of five tasks 1) writing a scrapper and crawling the Google results from 2) extracting features 3) clustering queries against the extracted features, 4) characterizing clusters and find out their representing keywords, 5) comparing the results with manually annotated intents. The contributions of our model are as follows:
\begin{itemize}
\item Our crawler is able to scrape Google's results without getting blocked.
\item It detects some intents which are different from the conventional intents.
\item It can automatically detect the intent of a given query using keywords extracted from the clustering.
\end{itemize}
This paper is organized to present related works in Section \ref{sec:SOTA}, methodology and feature extraction in Section \ref{sec:methodology}. We discuss the experiments in Section \ref{sec:experiments} and finally Section \ref{sec:conclusion} concludes this study, its findings and discusses future works.

\section{Related Works}
\label{sec:SOTA}
\textit{Intent Detection}: Intent detection generally is a classification problem aiming to find the intention of input data which is usually in the format of text. The models which are designed to identify intents base-on machine learning models can be divided into two groups \cite{zhou2017survey} in terms of exploiting hand-crafted features \cite{guo2010ready} \cite{jansen2007determining} versus embedding features \cite{pirvu2018predicting}. 
In the first group, researchers must extract some features which are important to identify query intent. Hand-crafted features can be derived from: i) Query tokens ii) Search Engine Results Pages (SERP) type and content tokens iii) interaction features (tracking user behavior including clicks log and queries log in a session).

In a study done by Sappelli \cite{sappelli2012collection}, a dataset of user queries is collected with the features of the topic, action type, expected result type (image, video, map, etc.), location sensitivity and so on. They studied the distributions of queries per each feature as well as the correlations between different features. Authors in \cite{kathuria2010classifying} took a transaction log from Dogpile into account to extract the required features such as query term, user id, time of day to train their clustering algorithm. They finally characterized the identified clusters as informational, transactional and navigational and demonstrated the most frequent words for each category. 

In a different study \cite{guo2010ready} Guo \etal investigated ﬁne-grained user interactions with the search results to identify user intent. According to this study, although considering all features together provides 97\% accuracy, features related to SERPs content are identified as the most important ones contributing to the classification.  Transactional queries are identified by CURL in \cite{sun2019clicked}. Authors in \cite{boteanusubjective2020} trained a classifier to label queries with the intents extracted from user reviews. In \cite{figueroa2015exploring} query-specific features, such as bag-of-words, length, recognized named entity, noun phrase, question and so on, are exploited to build three multi-class classifiers. 

In the second group, the models are using embedding features automatically derived from mainly neural networks. For example, a convolutional neural network model is designed in \cite{hashemi2016query} to extract the query embedding and use them to train the intent classifier. In an improved model of using word embedding, authors in \cite{sreelakshmi2018deep} proposed a deep learning-based platform using Bi-directional Long-Short Term Memory (BDLSTM). They have used word embedding from GloVe \cite{pennington2014glove} model and enriched them by bringing the synonyms and related words closer to each other in the vector space and moving the antonymous words away from each other. An automatic intent labeling model is introduced in \cite{pirvu2018predicting} using Recurrent and Convolutional Neural Networks (RNN, CNN). The model is trained against ground truth and some heuristic rules to perform a multi-intent prediction for unlabeled queries. One of the latest models for intent detection is Zero-shot User Intent Detection via Capsule Neural Networks \cite{xia2018zero}. This model considers new and not-seen intentions also. 

Two applications can benefit from intent detection researches, Search engines and SEO. Most of the mentioned researches that tried to detect the intent of a search query looking for solutions from the search engine perspective. Although SEOs can take advantage of those studies, dedicated research investigating the intent of queries from the SEO perspective is missing in the literature.
Our proposed model uses hand-crafted features extracted from SERPs to cluster queries for the sake of SEO. Although some studies investigated SERPs' correlation to the query to identify the query's intent, no research has studied this problem from the SEO perspective. 

\textit{Annotated Dataset}: To build and train an intent detection model, a manually labeled dataset is needed. The model learns how to identify the intention of new data after getting trained by labeled data. According to \cite{liu2019review}, there are several difficulties in the intent detection task. The most important one is the lack of annotated datasets. Labeling the intent of queries is usually done manually. Besides the dataset, the intention of the user is not always explicit. The ambiguity and implicit intention make this problem more complicated and difficult.

In \cite{pirvu2018predicting}, almost 2k queries are manually labeled including both test and train datasets. Later, authors automatically labeled the rest of their data by a classifier trained on the ground truth labeled dataset. A big dataset of 30k queries randomly selected from AOL web queries is manually annotated in \cite{figueroa2015exploring}. We use the last-mentioned dataset. We will discuss it in detail in Section \ref{subsec:dataset}.

\section{Methodology}
\label{sec:methodology}
Early, we reviewed the previous studies on detecting the intent of queries. However, our research is different from past studies due to the following reasons:
\begin{enumerate}
\item It targets the features that have never been investigated in the literature. 
\item Not only it will not rely on the manually annotated tags, (what most of the studies use them to design their classifiers) but also it will take advantage of a clustering model to verify the human-annotation of queries.
\item The Majority of the previous studies have addressed intent detection from the search engine's perspective. While this study looks at this problem from the perspective of SEO. Both are providing an automated method to identify the intent, but the second group provides additional advice to SEO to manage the content types which should be uploaded to the websites.
\end{enumerate}
This subsection describes in detail how we take advantage of Google's search results to build a model for intent detection.  
\subsection{Data Scrapper}
Data Scrapper is a python application, developed by our team, uses different techniques to send a query as a request to Google and Collects the results provided by Google. The script is written by Python and reads the queries from a public dataset provided in \cite{figueroa2015exploring} to request them from Google and collect the provided results. The major challenge of Scrapper is not getting blocked by Google. When Scrapper requests so many queries from Google, it is faced by a captcha. The captcha should be filled by a human. Therefore, our solution is using a proxy to request queries from different IPs. 

Scrapper collects different kinds of information from Google results shown in \ref{tbl:result_types}. This table shows each item with its description. To capture these features, Scrapper parses the HTML of the first page of the results. The results are saved in JSON files and then are processed to extract the final features. We will use that information in feature extraction and clustering processes. 
\begin{table}[!h]
\centering
\caption{Google's results types}
\begin{tabular}{|>{\centering\arraybackslash}m{2.7cm}|p{8.8cm}|}\hline
\textbf{Features} & \textbf{Description} \\ \hline
knowledge graph & It is a box that Google loads in the information related to each identified entity existing in the query from different sources.  \\\hline
Calculator  & It appears to answer directly the calculation-related queries.\\\hline
Direct answer & A box to respond a query that Google knows the answer. \\\hline
Map  & Direct answer to map related questions\\\hline
Local result &	It shows the possibility of local access to the searched term.\\\hline
Commercial-Sponsored & All the results showing the price. \\\hline
Twitter & If Google finds any tweets related to the searched term.\\\hline
Top Stories &	Google finds the recent news articles talking about the query. \\\hline
Videos & Very recent videos indicating the searched terms.\\\hline
Images & Categorized and recent images related to the query.  \\\hline
Content navigation bar	& Google provides a navigation bar of mainly objects such as movies, books on top of the search results.\\\hline
Featured Snippet &	It is a selected search result that answers the user's query right away. It can be a video, image, text, and so on.\\\hline
Rich Snippets &	Results in the form of cards having ratings and reviews. \\\hline
People Also Asked &	A list of Questions similar to the searched query\\\hline
Similar entity	& A list of related entities to the searched entity.\\\hline 
Google Translator	 & Representing the meaning or translation.\\\hline
Top-Button Ads & Links with ``Ad" next to their link	\\\hline 
Natural results & Natural blue links on the first page (organic links)\\\hline 
Partners block	& Links to Google's partners to search on their websites \\\hline 
Other cards & 	Boxes similar to the twitter block, such as popular products. \\\hline 
\end{tabular}
\label{tbl:result_types}
\vspace{-0.4cm}
\end{table}
\subsection{Feature Extraction}
Feature extraction is done after collecting Google results. We searched in the provided results for different types of results such as images, videos, featured snippets, rich snippets, knowledge graphs, direct answers, ``people also ask'' and so on. We consider each of those result types as a feature and extracted their title, number, and position. The idea behind this is to find what kind of information Google recognizes to show for each keyword. As the aim of SEO is to increase the visibility of a website and consequently a brand, thus, SEO consultants could easily decide what kind of content should be presented in the clients' website to get easily visible by Google. Understanding Google's methodology will lead to providing greater consultation for related and proper content on the website.

As mentioned, features are extracted from different types of Google results. Our goal is to identify some groups of queries and consequently some keywords for which Google expects the websites to provide content in special formats.

\subsection{Clustering }
After feature extraction, now we need to apply a clustering model to find Informational, Transactional, and Navigational groups of keywords. As we pointed out before, there is a possibility that we end up in different categories than those three. To do clustering, we choose the KMeans algorithm \cite{macqueen1967some}. The most proper number of clusters is found by the Elbow method to be 3. 

\subsection{Datasets}
\label{subsec:dataset}
As we discussed in the state-of-the-art section, almost all the researches in this area use manually labeled data. Due to using clustering technique, our method does not need any labeled data. However, we use the labeled dataset to build an opportunity to compare the clustering results against the human-labeled tags. We use the public labeled dataset introduced in \cite{figueroa2015exploring}. Table \ref{tbl:dataset} shows the characteristics of the dataset used in our study. Majority of the queries in the selected-AOL dataset are informational.
\begin{table}[!h]
\vspace{-0.7cm}
\centering
\caption{The Dataset Characteristics}
\begin{tabular}{|c|c|c|c|c|} \hline 
\textbf{Dataset name} & \textbf{\#Queries} & \multicolumn{3}{c|}{\textbf{Manual-Labels}}\\ \cline{3-5}
& & Informational & Navigational & Transactional \\ \hline 
Selected-AOL \cite{figueroa2015exploring} & 30k	& 23700 &	4574	& 1678 \\ \hline
\end{tabular}
\label{tbl:dataset}
\vspace{-0.5cm}
\end{table}

The Scrapper application crawls the public dataset's queries from Google and saves the results. It is worth mentioning that Google's SERPs are very user-dependent and its results are different from user to user. To have organic results, we run the scrapper on a server with neither search history nor logged in user. After crawling, we divided the dataset into the train and test sets with 90\% and 10\% population, respectively.

\section{Clustering Experiments}
\label{sec:experiments}
\begin{figure}[!h]
\vspace{-0.5cm}
\begin{subfigure}{.5\textwidth}
\includegraphics[scale=0.3]{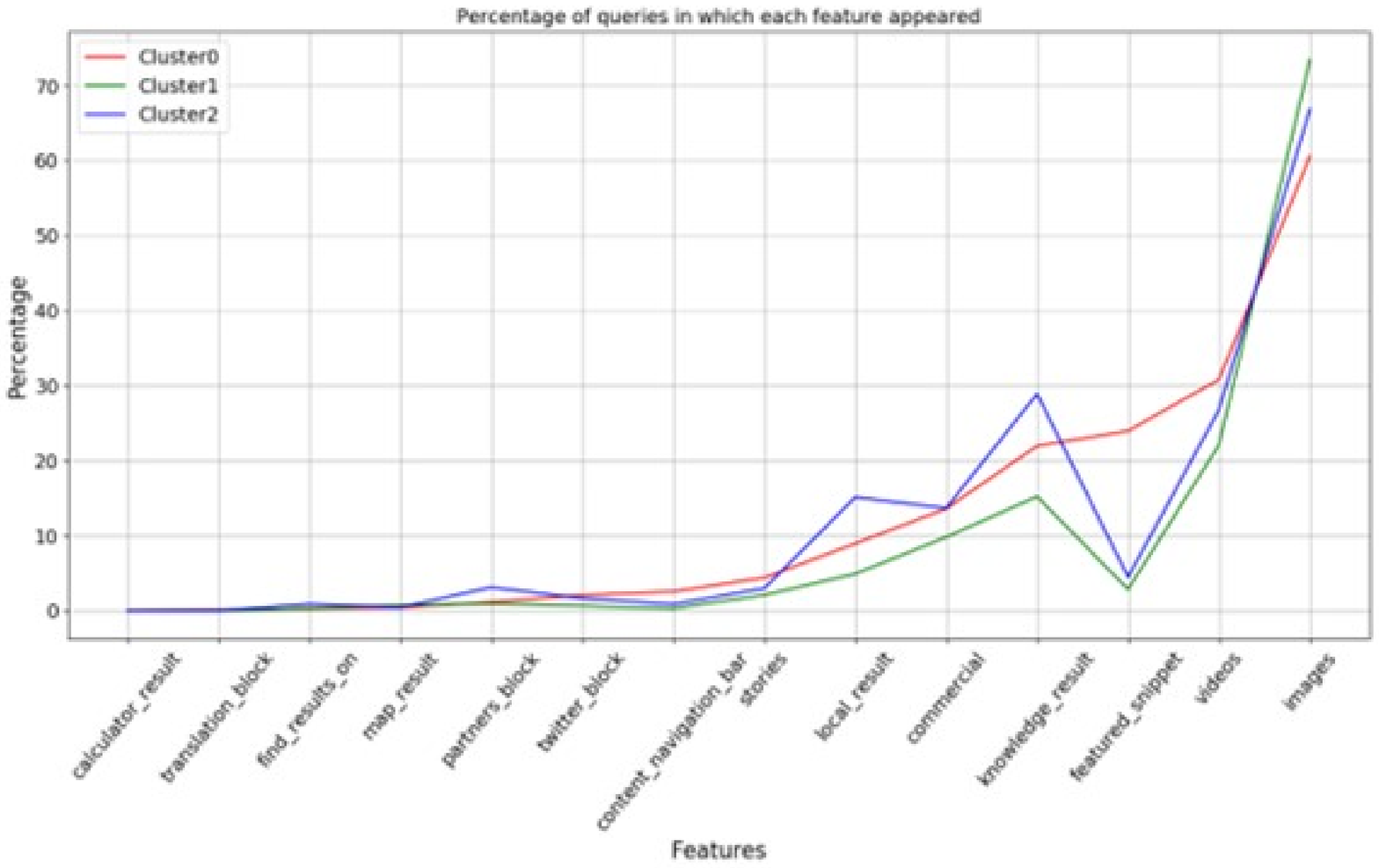}
\caption{Binary features value for each cluster}
\label{fig:binary}
\end{subfigure}
\begin{subfigure}{.5\textwidth}
\includegraphics[scale=0.38]{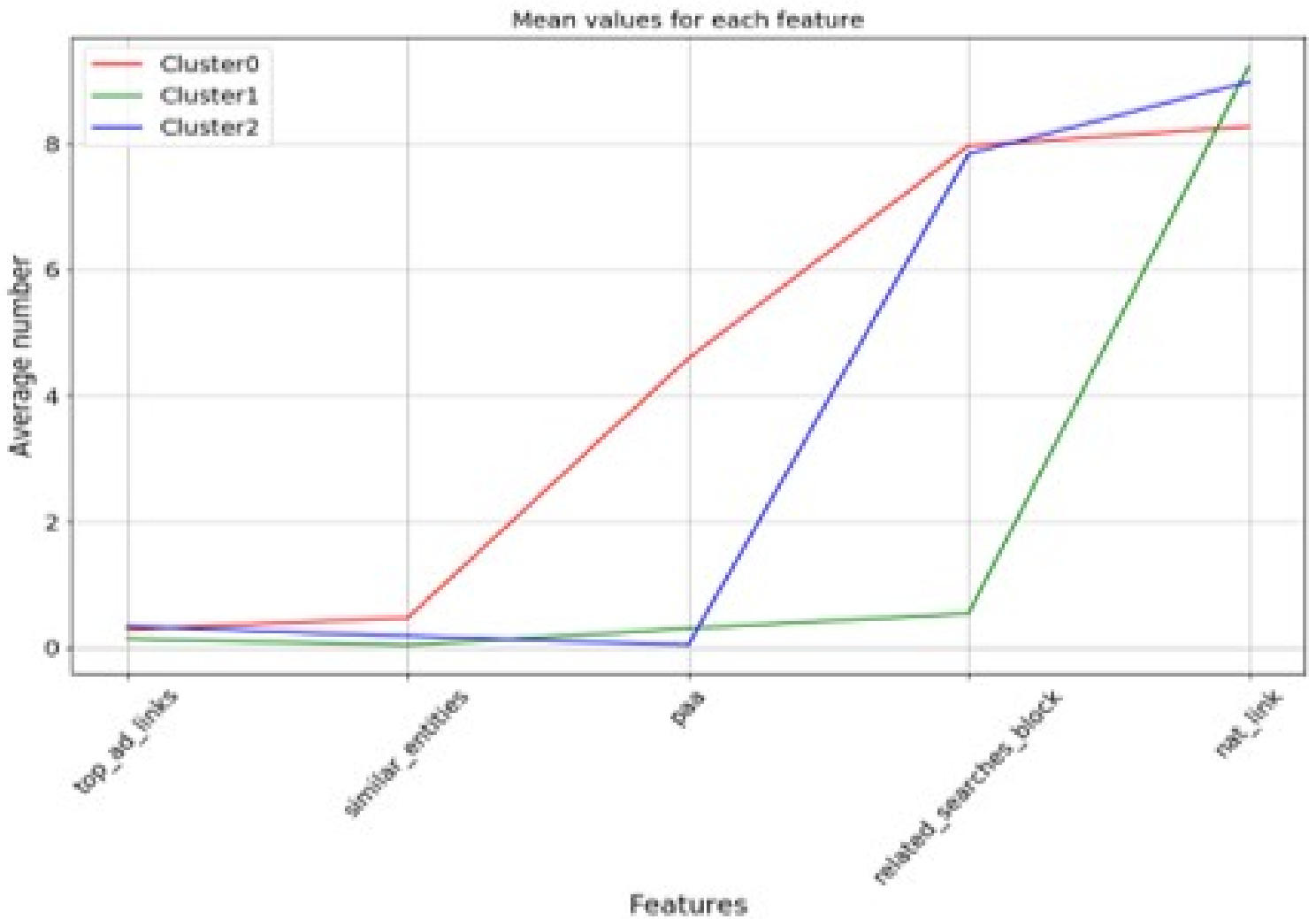}
\caption{Numeric features value for each cluster}
\label{fig:numeric}
\end{subfigure}
\caption{Features values}
\label{fig:features}
\vspace{-0.5cm}
\end{figure}

Before running clustering, we investigated the correlation of features to exclude the tightly correlated ones. Due to not founded any correlated features, we run the KMeans model with 19 features and K=3. Characterizing clusters is ended up with very interesting clusters which are different from three predefined classes (including \textit{Informational, Navigational}, and \textit{Transactional}). Table \ref{tbl:clusters} shows the distribution of queries and their tags in clusters. 
\begin{table}[!h]
\vspace{-0.7cm}
\centering
\caption{Distribution of queries into clusters}
\begin{tabular}{|c|c|c|c|c|} \hline
\textbf{Cluster Name}& \textbf{\#Queries} & \textbf{Informational} & \textbf{Transactional}& \textbf{Navigational} \\ \hline
Cluster0 & 	11582	& 9581 & 	530 & 	1471 \\ \hline
Cluster1 & 	5771 & 	4392	& 472 & 	907 \\ \hline
Cluster2	&  9603 & 	7339	& 529 & 	1735  \\\hline
\end{tabular}
\label{tbl:clusters}
\vspace{-0.5cm}
\end{table}
As an initial step to characterize the clusters, we study the value of the features for each cluster. We divided features into two groups, features with binary and numeric values. Figure \ref{fig:binary} and \ref{fig:numeric} show the binary and numeric feature values for each cluster, respectively. 

We first go through the binary features' plot. Each value in the plot refers to the percentage of the True values of each feature. 
\begin{itemize}
\item	Cluster0 has a noticeable higher value of featured snippets as well as a  slightly higher value in the navigation bar feature. While the other two clusters have a very small value of featured snippet. It can be a piece of initial evidence for cluster0 of being information seeking queries. 
\item	The only feature that has a higher value in cluster1 is \textit{images}!
\item	Where two clusters 0 and 2 have almost the same values for commercial, this value for cluster1 is noticeably low. It shows low relation of queries in cluster1 with shopping intent.
\item	Cluster2 has a significantly high value of local results, knowledge results and somehow partners block, which is more likely to include queries having \textit{Entities}\footnote{The Google's Knowledge Graph has millions of entries that describe real-world entities like people, places, and things. These entities form the nodes of the graph, and are called Knowledge Graph Entities \cite{supportGoogle}} identified by Google and local information. Accompanying these two types of results can be interpreted as the queries that are looking for local special places. Later, looking at the words of queries will give more information about their exact intent.
\end{itemize} 

In the numeric features' plot, the values indicate the mean value of features for each cluster. It also shows interesting points:
\begin{enumerate}
\item	While the mean value of PAA is near to 0 for cluster1 and cluster2, cluster0's queries have an average value of 5. This observation strengthens the probability of cluster0 to be an informational cluster of queries. 
\item	In Figure \ref{fig:numeric}, the related searches value for cluster1 is almost 0 which is strange! While for the other two clusters is almost 8 which is the regular number of suggestions by Google. To discover the reason, we have to look at the vocabulary of the queries in this cluster.
\end{enumerate}

\begin{figure}[!h]
\begin{subfigure}{.3\textwidth}
\centering
\includegraphics[scale=0.2]{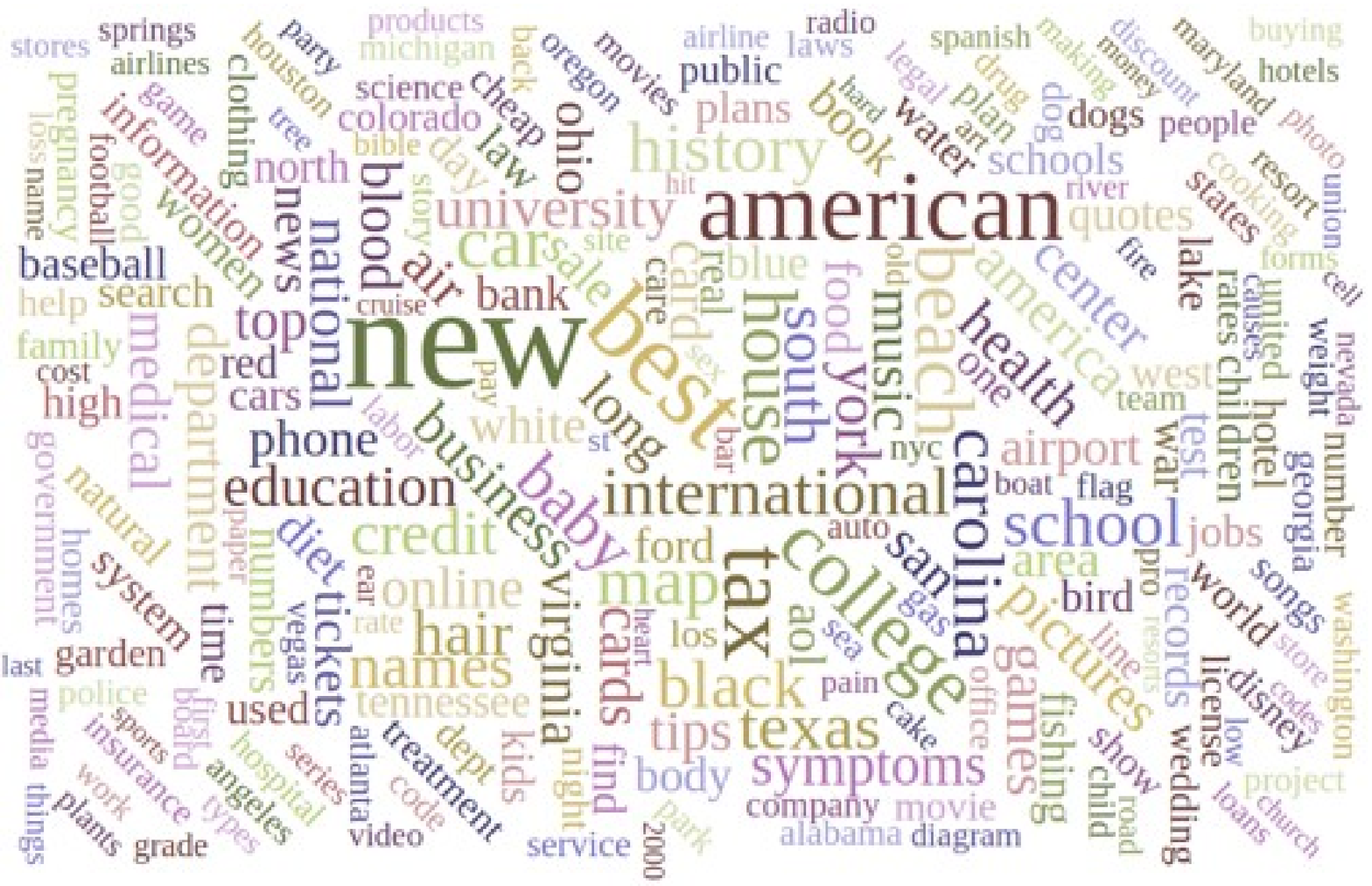}
\caption{Cluster0}
\label{fig:wc_c0}
\end{subfigure}
\begin{subfigure}{.3\textwidth}
\centering
\includegraphics[scale=0.2]{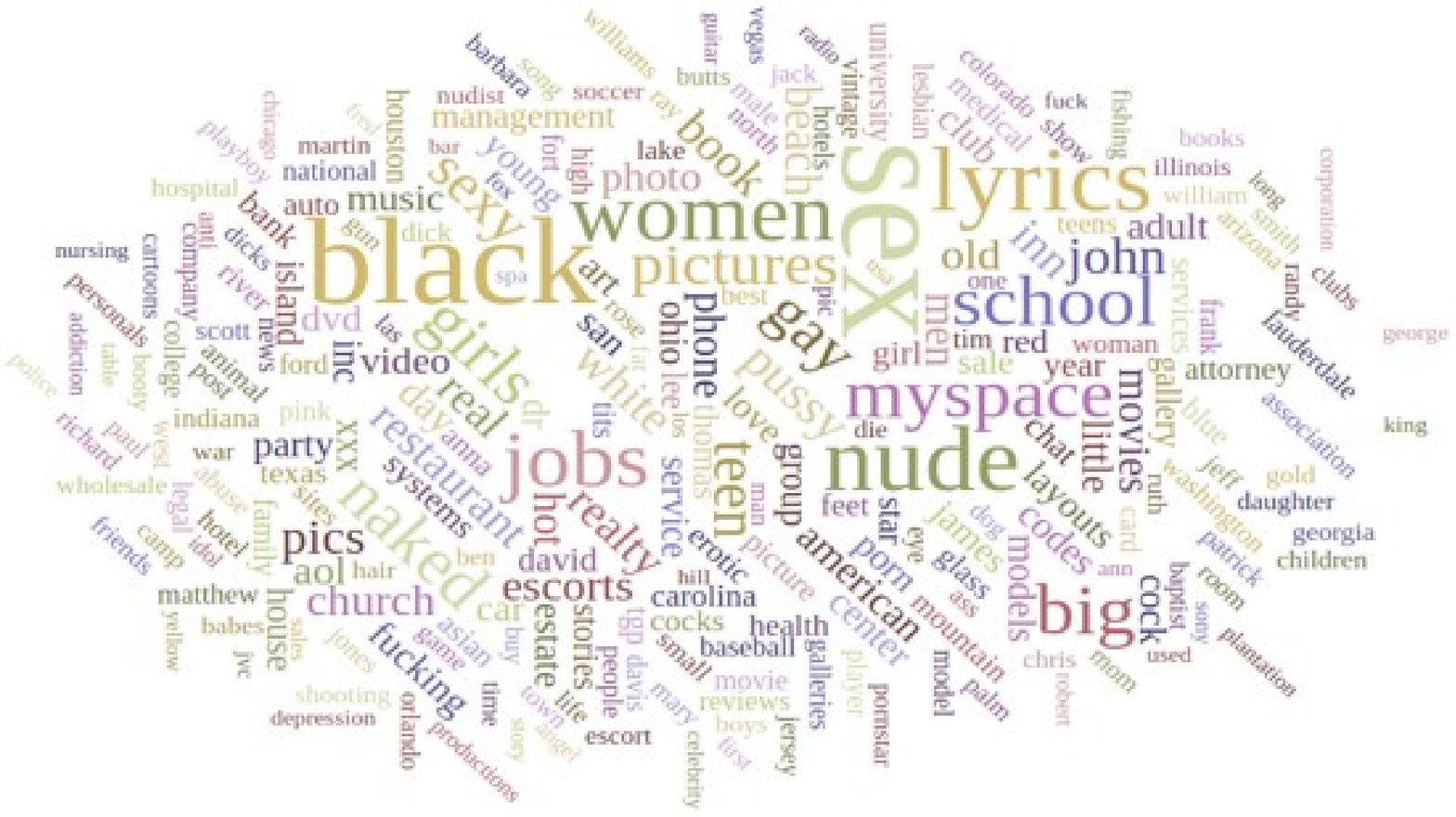}
\caption{Cluster1}
\label{fig:wc_c1}
\end{subfigure}
\begin{subfigure}{.34\textwidth}
\centering
\includegraphics[scale=0.2]{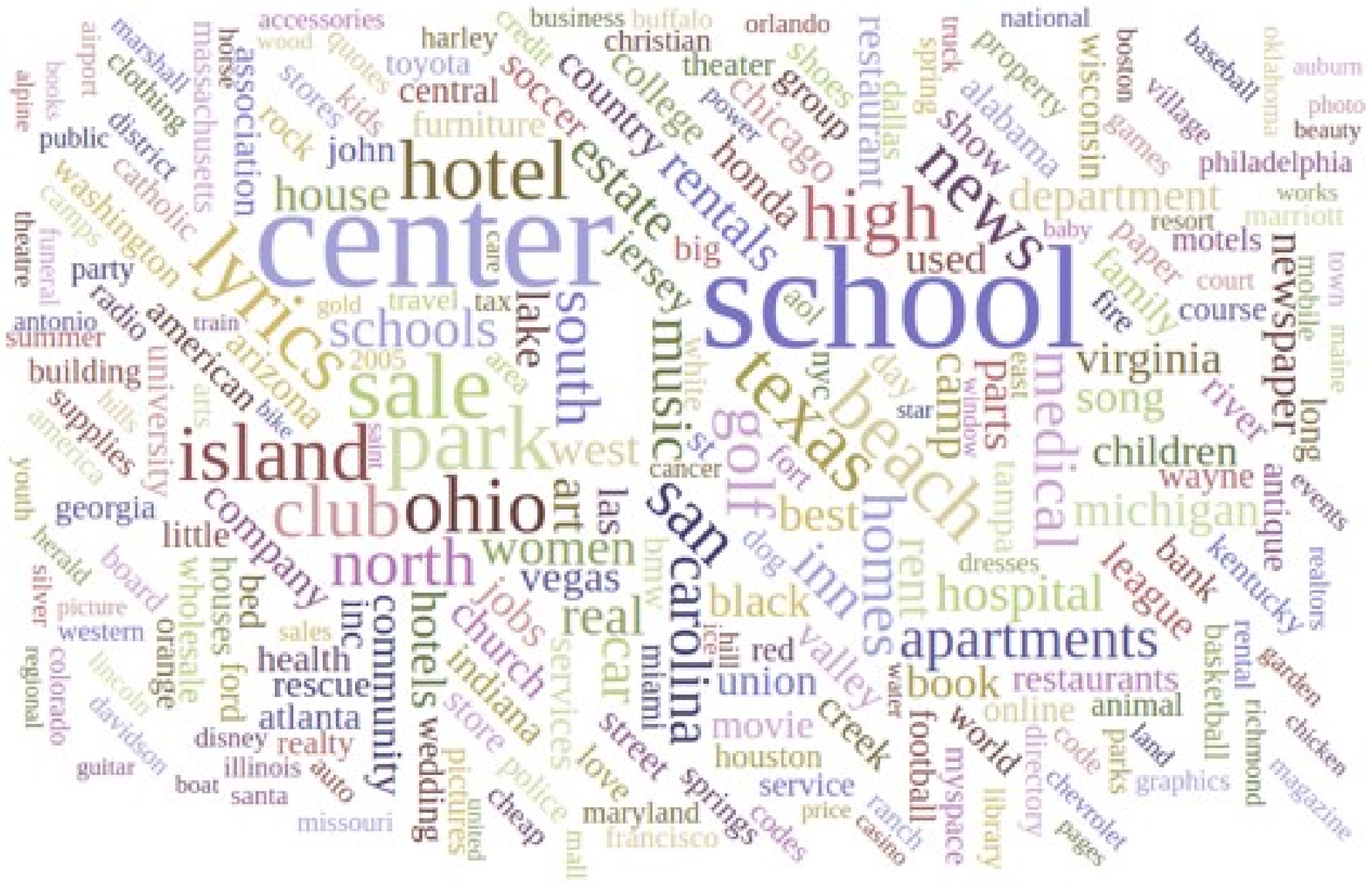}
\caption{Cluster2}
\label{fig:wc_c2}
\end{subfigure}
\caption{Wordcloud distribution per cluster}
\end{figure}

So far, from the observations of the features' values, we found cluster0 to include more informational (questions with a direct answer), and cluster2 more local queries. 
To find out more about the clusters, we plot the distribution of the words\footnote{https://www.jasondavies.com/wordcloud/} of the queries in each cluster. 
 In Figure \ref{fig:wc_c0}, the bigger size of the vocabularies indicates more repetition. Big words such as \{\textit{new, best, americans}\} shows that they get repeated more than other words in the queries of this cluster. Looking at the other less big words such as \{\textit{tax, business, car, education, health, house, college, university}\} and putting them besides the most frequent ones lead to a representation of queries which mainly are searched to acquire information. Relying on the results so far discovers that Google tries to show featured snippet and PAA for this kind of general informational queries.
In Figure \ref{fig:wc_c1}, it is observable that the most frequent words are \{\textit{black, sex, women, nude, lyrics}\}. Putting these words next to the other words in this group generates some sexual or racist phrases. The most highlighted feature of this cluster in our analysis from the previous section is almost zero number of keywords related to the query. The zero number of this feature for the queries of cluster1 illustrates that Google may not show similar keywords and queries to its users when their query carries a sexual or racist intent to may stop users from searching these kinds of queries.

Finally, in Figure \ref{fig:wc_c2}, the most frequent vocabularies are \{\textit{center, school, park, beach, island, club, hotel, sale}\}. Almost all of these words refer to locations. Based on our feature analysis, the dominant features of this cluster are \textit{local result}, \textit{knowledge graph} and \textit{partners block}. The observations convey that cluster2 is mainly a collection of queries that are looking for local information including places (such as schools, hotels, islands, beaches, parks, and so on). 

Our model identified some intents (consisting of general qualitative information, racist/sexual intent, and local/place information) different than the conventional intents. The conventional intents are important for search engines. While the intents identified by our model is mainly for SEO. In a nutshell, the method is concluded to three clusters, i) Cluster0: general informative and qualitative queries, ii) Cluster1: queries with sexual and racist intent, iii) Cluster2: local and places information.  

Although, all the queries grouped in different clusters are not exactly from the same context, the wordcloud representation of clusters can help us to extract some keywords to automatically tag the intent of a given query to be in one of the above clusters. 

We processed the frequent words to exclude common, unrelated, and ambiguous words for each intent. To test the functionality of the extracted keywords, we automatically labeled the intent of the test queries based on their words. The test set with a 10\% population is used for this experiment. In case of an equal number of words, we consider the maximum priority for Informational and the minimum priority for Sexual/Racism intents. 
From the other side, the queries inside the test dataset are scrapped from Google and the designed clustering model is applied on this dataset. The labels out of the clustering are compared to the labels out of the automatic labeling (vocabularies-filtering).  Table 6 shows the results. Total number of queries is almost 3k with 2.4k \textit{Informational}, 460 \textit{Navigational}, and 150 \textit{Transactional} intents. 

\begin{table}[!h] 
\vspace{-0.7cm}
\centering
\small
\caption{Clustering vs. vocabulary-based intent tagging results}
\begin{tabular} {|>{\centering\arraybackslash}m{1.5 cm}| >{\centering\arraybackslash}m{2.1cm}| >{\centering\arraybackslash}m{1.8cm}| >{\centering\arraybackslash}m{1.7cm}| >{\centering\arraybackslash}m{2.0cm}| >{\centering\arraybackslash}m{1.2cm}| >{\centering\arraybackslash}m{0.9cm}|} \cline{3-5}
\multicolumn{2}{c}{} & \multicolumn{3}{|c|}{\scriptsize{vocabulary-based intents (predicted)}} & \multicolumn{2}{c}{} \\ \cline{3-7}
\multicolumn{2}{c|}{} & \scriptsize{Informational} &	\scriptsize{Local/Place  information}	 & \scriptsize{Sexual/Racism} 	& \scriptsize{Precision}	& \scriptsize{Recall} \\  \hline
\multirow{3}{1.5cm}{\scriptsize{Clustering intents (Actual)} } & \scriptsize{Informational}	& 1232 & 	54 & 	25	& 0.46\%	& \textbf{0.94\%}\\ \cline{2-7}
 & \scriptsize{Local/Place  information}	& 904	& 141 & 	26	 & \textbf{0.64\%}	 & 0.13\% \\  \cline{2-7}
& \scriptsize{Sexual/Racism}	& 	519	& 	25	& 	70	& 	\textbf{0.58\%}	& 	0.11\% \\ \hline
\end{tabular}
\label{tbl:tagging}
\vspace{-0.5cm}
\end{table}

In Table \ref{tbl:tagging}, the labels on the left side and the top are the outputs of clustering and vocabulary-based labeling, respectively. We consider the outputs of the clustering as the \textit{actual labels} and compare them with the outputs of the vocabulary-based labeling as the \textit{predicted labels}. Recall indicates the percentage of the queries in each intent which are predicted correctly inside that intent. According to the results, 94\% of informational queries are correctly predicted as informational. While in the other intents, the low percentage of recall shows that most of their queries are assigned to some intents different than their intent. Two possible reasons can cause this low percentage, low accuracy of clustering and low efficiency of the extracted keywords.
As we can not manually check the intent of each query, we will not be able to judge the accuracy of the clustering. While investigating the second reason, we computed the precision values. High precision values for Local information and Sexual/Racism clusters indicate the efficiency of the keywords on identifying the right queries for those two intents. It means that the keywords are chosen precisely in such a way that they hardly misidentify the queries (with other intents) inside these two intents. 

As a result, as we mentioned earlier, although some frequent keywords are representing each cluster, there remain some queries in each cluster that have none of their words in the representative keywords. Those queries are mislabeled in the Informational cluster using the vocabulary-based labeling.  
Consequently, the vocabulary-based method has a limitation which constraints it to label only the queries in which there are at least one of the selected keywords. This limitation persuades us to strengthen our model by taking other methods into account. As future work, we will take advantage of manual labels and Google's BERT model to identify the intents.

\section{Conclusion}
\label{sec:conclusion}
In this study, we have proposed and developed a model to identify the intent of user queries. The presented approach has the novelty of i) crawling Google SERPs results, ii) using new features (which have never been studied before), iii) identifying new and more detailed intents, iv) and finally, studying intent detection problem from the SEO perspective. Although Google has previously provided very few words for each identified intents, without any solid method to extend those words, our clustering model can provide three sets of keywords to automatically identify the query's intent. 

As a future work, we will study the semantic and NLP relation between the query and those features that have text, such as featured snippets, PAA, and knowledge graph.
We will use as well the fine-tuned BERT model to identify the intention of user queries and compare the results with the baseline methods.

\bibliographystyle{IEEEtran}
\bibliography{citations}

\end{document}